\documentstyle[epsfig,aps]{revtex}
\begin{document}

\title{Effect    of    transients    in    nuclear   fission   on
multiplicity of prescission
neutrons }

\author{ Gargi Chaudhuri \thanks{Electronic address:
gargi@veccal.ernet.in} and Santanu Pal\thanks{Electronic address:
santanu@veccal.ernet.in}}

\address{Variable  Energy  Cyclotron  Centre,  1/AF Bidhan Nagar,
Kolkata 700 064, India}
\maketitle
\begin{abstract}
Transients in the fission of highly excited nuclei are studied in
the  framework  of  the Langevin equation. Time-dependent fission
widths are calculated which show that a steady flow  towards  the
scission point is established,  after the initial transients, not
only for nuclei which have  fission  barriers but also for nuclei
which have no fission barrier. It is shown from a  comparison  of
the transient time and the fission life time that fission changes
from  a diffusive to a transient dominated process over a certain
transition region as a function of the  spin  of  the  fissioning
nucleus. Multiplicities of prescission neutrons are calculated in
a  statistical  model  with  as  well  as  without a single swoop
description of fission and  they  are  found  to  differ  in  the
transition  region.  We  however  find  that  the  difference  is
marginal and hence a single swoop picture of fission  though  not
strictly  valid in the transition region can still be used in the
statistical model calculations.

\end{abstract}

\pacs{  PACS numbers: 05.40.Jc Brownian motion, 24.60.Lz Chaos in
nuclear systems, 25.70.Jj Fusion and fusion-fission reactions}

\eject

\noindent  {\Large {\bf 1 Introduction}}
\\

The   fission  dynamics  of hot compound nuclei continues to be a
subject of considerable interest essentially due to the fact that
the fission life time determined from the phase space argument of
Bohr and Wheeler turns out to be  too  small  to  allow  for  the
rather  large  number  of experimentally observed light particles
evaporated prior to fission \cite{thoe}. It  was  therefore  felt
necessary to look beyond the statistical model and this gave rise
to  a  revival  of  interest  in the original work of Kramers who
considered fission as  a  diffusive  probability  flow  over  the
fission  barrier  \cite{kram}.  Dissipative  dynamical  models of
fission were subsequently developed to include particle  emission
and   were  found  to  reproduce  a  host  of  experimental  data
\cite{gran1,ab1,fro1}.

In  a  dissipative  dynamical model, the dynamics associated with
the fission degree of freedom is usually considered to be similar
to  that  of a Brownian particle floating in a viscous heat bath.
The heat bath in this picture represents  the  rest  of  all  the
other  nuclear  degrees  of  freedom  which  are assumed to be in
thermal  equilibrium.  The interaction between these large number
of intrinsic degrees of freedom and the fission degree of freedom
gives  rise to a random force and consequently a dissipative drag
on the dynamics of fission \cite{ab1}. The fission trajectory can
thus be followed by solving  the  appropriate  Langevin  equation
\cite{ab1,fro1}.  Alternatively, one can study the time evolution
of the phase space density of the fission coordinates  using  the
Fokker-Planck equation \cite{gran1,gran2}.

Several  time  scales  are of physical significance as a compound
nucleus makes its journey from its  formation  to  its  scission.
Considering  an  ensemble of compound nuclei which are all formed
at a given instant, it requires a certain  interval  of  time  in
order  to  develop  a steady probability flow at the saddle point
across the fission  barrier.  During  this  time  interval,  also
referred  to  as  the transient time ($\tau$) \cite{gran3,gran4},
the probability  flow  at the saddle point increases from zero to
its statioanry  value.  This  stationary  probability  flow  also
defines  the  stationary  fission  width  ($\Gamma_{0}$)  and the
associated fission life  time  ($\tau_  {f}=  \hbar/\Gamma_{0}$).
However,  the  entire  fission process itself becomes a transient
when the compound nucleus has no fission barrier. In such  cases,
it  is  usually  argued  that  the  full distribution reaches the
scission point in  a  single  swoop  \cite{gran3}.  The  compound
nucleus  then  survives  exactly  the  swooping  down time ($\tau
_{s}$) in this picture. These transients in nuclear fission  were
considered  earlier  in great details in a series of publications
\cite{gran3,gran4,wm1}.   Using   a   number    of    simplifying
assumptions,  analytical  expressions  were obtained for the time
dependent fission widths for both the weakly damped and  strongly
damped  motions  \cite{gran4,wm1}.  The  stationary  value of the
fission width was obtained by Kramers  earlier  \cite{kram}.  The
transient  time  for  cases  where  there  is no barrier was also
obtained \cite{gran3}.

An  highly  excited  nucleus can emit a number of light particles
(including photons) before it  undergoes  fission.  In  order  to
trace  the  time  evolution  of  such  a  hot  nucleus,  it would
therefore be most desirable to couple a dynamic theory of fission
with the statistical emission of light particles. Extensive  work
has been done in this direction using either the Fokker-Planck or
the  Langevin  equations and the importance of the dynamic nature
of the fission process at high excitations has  been  established
\cite{gran1,ab1,fro1}.   Such   calculations  however  are  quite
involved and require a large computer time.  An  alternative  and
easier   approach   would  be  to  perform  a  statistical  model
calculation by modifying a  cascade  code  in  which  fission  is
treated  as  one of the decay channels and the time dependence of
the fission width is explicitly taken into account \cite{butsch}.
In such calculations, the input fission widths and the  transient
times   are   usually   taken  from  the  analytical  expressions
\cite{kram,gran1,gran2}.

In  the  present work, we would examine certain issues related to
the time dependence of fission  widths  and  its  effect  on  the
multiplicity  of  the prescission neutrons. First, we would study
the effect of lowering the fission barrier on the time dependence
of the rate of fission. We would use  the  Langevin  equation  to
model  the  fission  dynamics for our purpose. The motivation for
this  study is to find the transition from a diffusive process in
the presence of  a  fission  barrier  to  a  transient  dominated
picture  when  there  is no fission barrier. We would indeed find
that the diffusive nature of fission  continues  to  some  extent
even  for  cases  which  have  no fission barrier. The underlying
physical picture that would emerge for fission in the absence  of
a  fission  barrier  would be as follows. Consider an ensemble of
fission trajectories which have started together sliding down the
potential (with no fission barrier) towards the  scission  point.
However,  the  random  force  acting  on  the  trajectories  will
introduce a dispersion in their arrival  time   at  the  scission
point.  In  other words, the trajectories will cross the scission
point at different instants and a flow will thus  be  established
at  the  scission  point.  However, the effect of this dispersion
will be reduced when the conservative force becomes much stronger
than the random force. This would happen at  very  large  angular
momentum  of the fissioning nucleus due to the strong centrifugal
force. Therefore, the single swoop picture  for  fission  becomes
more  appropriate at very large values of spin of the nucleus. We
would establish the above scenario in the first part of our work.

Next,   we  would  perform  statistical  model  calculations  for
prescission neutron multiplicity using the time dependent fission
widths  as  well as the single swoop description of fission. This
would be done with the aim of finding how  well  the  statistical
model  calculations  with and without the single swoop assumption
agree with  each  other.  We  would  subsequently  calculate  the
prescission  neutron multiplicity in a dynamical model of fission
and would compare  the  results  with  those  obtained  from  the
statistical  calculations.  Though  one  would expect the results
from the statistical and the dynamical  calculations  to  be  the
same,  there could be some differences due to reasons such as the
following one. The neutron width at any instant of time evolution
of a hot nucleus depends upon its temperature. However, a part of
the total excitation energy is used to build  up  the  collective
kinetic  energy  in  the  fission channel in the dynamical model.
This reduces the available intrinsic excitation energy, and hence
the  temperature,  in  the  dynamical  model  compared   to   the
statistical  model  calculation where the total excitation energy
is taken as the intrinsic excitation energy of the nucleus.  This
could  reduce  the neutron multiplicity in a dynamical model than
that obtained from a statistical calculation. We would  ascertain
the magnitude of such differences from our calculation.

The  time  dependent  fission widths will be evaluated by solving
the Langevin equation numerically in the present paper.  We  have
to  resort  to  numerical  methods  rather  than using analytical
solutions  due to the fact that  analytical  solutions  assume  a
constant friction whereas we would use a strongly shape dependent
friction  force in our calculation. This shape dependent friction
is  a  modified  version  of  the  one-body  wall  friction,  the
so-called  "chaos-weighted  wall formula", in which the effect of
chaos in the single-particle motion is incorporated  \cite{pal1}.
The  prescission  neutron  multiplicity  and  fission probability
calculated from Langevin dynamics using the  chaos-weighted  wall
friction  were  found  to agree fairly well with the experimental
data for a number of heavy compound nuclei ($A \sim 200$) over  a
wide   range   of  excitation  energies  \cite{pal3}.  Since  the
chaos-weighted wall friction contains no adjustable parameter  to
fit  the  experimental  data,  we  consider  this  to be the most
appropriate dissipative force to describe the dynamics of fission
and accordingly we shall use it here. Further,  we  have  already
reported a systematic study of fission widths using this friction
\cite{pal2}.  This  study  was  confined  to  cases  with fission
barriers whereas we would concentrate upon fission in the absence
of a barrier in the present work.

We  shall  give  a  brief  description of the dynamical model for
fission in the next section. Section 3 will contain the numerical
results. A summary of the results along with the conclusions will
be presented in the last section.
\\
\\
\\
\noindent {\Large {\bf 2 Langevin dynamics of fission}}
\\
\\
\noindent{\large {\bf 2.1 Nuclear shape, potential and inertia}}
\\
\\
We  shall  choose  the  shape  parameters  $c,h$  and $\alpha$ as
suggested by Brack {\it et al.} \cite{brack}  as  the  collective
coordinates  for  the fission degree of freedom. However, we will
simplify the calculation by considering  only  symmetric  fission
($\alpha=0$).  We  shall  further assume in the present work that
fission would proceed along the valley of the potential landscape
in $(c,h)$ coordinates though  we  shall  consider  the  Langevin
equation  in  elongation  $(c)$  coordinate  alone  in  order  to
simplify  the  computation.  Consequently,  the   one-dimensional
potential   in   the   Langevin   equation  will  be  defined  as
$V(c)=V(c,h)$ {\it at valley} and  the coupled Langevin equations
in one dimension will be given \cite{ab2} as

\begin{eqnarray}
\frac{dp}{dt}   &=&
-\frac{p^2}{2}   \frac{\partial}{\partial   c}\left({1  \over  m}
\right) -
   \frac{\partial F}{\partial c} - \eta \dot c + R(t), \nonumber\\
\frac{dc}{dt} &=& \frac{p}{m} .
\label{(1)}
\end{eqnarray}

\noindent  The  shape-dependent  collective  inertia    and   the
friction coefficient in the above equations are  denoted  by  $m$
and $\eta$ respectively. The free energy of the system is denoted
by $F$ while $R(t)$ represents the random part of the interaction
between the fission degree of freedom and the rest of the nuclear
degrees  of  freedom considered collectively as a thermal bath in
the  present  picture.  The  collective  inertia,  $m$,  will  be
obtained  by  assuming  an  incompressible  irrotational flow and
making  the  Werner-Wheeler  approximation   \cite{davies}.   The
driving  force  in  a thermodynamic system should be derived from
its free energy which we will calculate considering  the  nucleus
as  a  noninteracting  Fermi  gas  \cite{fro3}. The instantaneous
random force $R(t)$ is modeled after that of a  typical  Brownian
motion and is assumed to have a stochastic nature with a Gaussian
distribution  whose  average  is zero \cite{ab1}. The strength of
the random force will be determined by the  friction  coefficient
through   the   fluctuation-dissipation   theorem.   A   detailed
description of the various input quantities may  be  found  in  a
recent publication \cite{pal2}.
\\
\\
\\
\noindent{\large {\bf 2.2 Nuclear friction}}
\\
\\
In  order  to  define the friction force in the Langevin equation
(eq.(1)), we will consider the  one-body  wall and window
dissipation  \cite{blocki}.  We  will use the chaos-weighted wall
friction as the one-body wall dissipation. We shall give  here  a
brief  description  of  the  chaos-weighted  wall  friction,  the
details  of   which   is   given   elsewhere   \cite{pal1}.   The
chaos-weighted friction coefficient ${\eta_{cwwf}}$ is given as

\begin{eqnarray}
\label{fric}
\eta_{cwwf} = \mu \eta_{wf},
\label{(2)}
\end{eqnarray}

\noindent  where ${\eta_{wf}}$ is the friction coefficient as was
given by the original wall formula \cite{blocki} and ${\mu}$ is a
measure of chaos (chaoticity) in the  single-particle  motion  of
the  nucleons  within  the  nuclear  volume  and  depends  on the
instantaneous shape of the nucleus \cite{pal1}.  In  the  present
classical  picture, this will be given as the average fraction of
the nucleon trajectories that are chaotic when  the  sampling  is
done  uniformly over the nuclear surface. The value of chaoticity
is evaluated for a given shape of the nucleus by sampling over  a
large   number  of  classical  nucleon  trajectories  while  each
trajectory is identified either as a regular or as a chaotic  one
by  considering  the  magnitude  of its Lyapunov exponent and the
nature of its variation with time \cite{blocki2}. It is  observed
that  the  value  of  chaoticity $\mu$ changes from 0 to 1 as the
nucleus evolves from a spherical shape to a highly deformed  one.
This  implies  that  the  wall  friction  is  very small for near
spherical shapes of a nucleus. The physical picture  behind  this
is  as  follows.  A  particle  moving  in  a spherical mean field
represents a  typical  integrable  system  and  its  dynamics  is
completely  regular.  When  the boundary of the mean field is set
into motion (as in fission), the energy gained by the particle at
one instant as a result of a collision with the  moving  boundary
is  eventually  fed  back to the boundary motion in the course of
later collisions \cite{koo}. An integrable  system  thus  becomes
completely  nondissipative  resulting  in  a  vanishing  friction
coefficient. On  the  other  hand,  the  nucleon  motion  becomes
chaotic  in  a  highly deformed nucleus. Consequently, the energy
tranfer  from  boundary  motion  to   particle   motion   becomes
irreversible  giving rise to a large friction force acting on the
boundary motion. These  aspects  were  established  from  various
numerical   investigations   \cite{schuck,blocki3}   and   was
incorporated into the wall formula  to  give  the  chaos-weighted
wall friction \cite{pal1}.

In  the wall and window model of one-body dissipation, the window
friction is expected to be effective after a neck  is  formed  in
the nuclear system \cite{sierk2}. Further, the radius of the neck
connecting the two future fragments should be sufficiently narrow
in  order  to  enable a particle that has crossed the window from
one side to the other to remain within the other fragment  for  a
sufficiently  long  time. This is necessary to allow the particle
to undergo a sufficient number of  collisions  within  the  other
side  and  make  the  energy  transfer irreversible. It therefore
appears that the window friction should be very nominal when neck
formation just begins. Its strength should increase as  the  neck
becomes  narrower  reaching  its  classical  value  when the neck
radius becomes  much  smaller  than  the  typical  radii  of  the
fragments.  We  however  know  very little regarding the detailed
nature of such a transition.  We  shall  therefore  refrain  from
making  any  further  assumption  regarding  the  onset of window
friction. Instead, we shall define  a  transition  point  in  the
elongation  coordinate $c_{win}$ beyond which the window friction
will be switched on. We  shall  also  assume  that  the  compound
nucleus  evolves  into  a  binary  system  beyond  $c_{win}$  and
accordingly correction terms for the motions of  the  centers  of
mass  of  the  two halves will be applied to the wall formula for
$c>c_{win}$ \cite{sierk2}. However, it may be  noted  that  while
the  window  friction  makes a positive contribution to the total
friction, the centre of mass motion correction reduces  the  wall
friction.  The  two  contributions therefore cancel each other to
some extent. Consequently, the resulting wall-and-window friction
is not very sensitive to the choice of the transition  point.  We
shall  choose  a  value  for $c_{win}$ at which the nucleus has a
binary shape and the neck radius is half of the radius of  either
of the would-be fragments.

The chaos weighted wall and window friction will thus be given as

\begin{eqnarray}
\eta(c < c_{win})=  \eta_{cwwf}(c < c_{win}),
\label{(3)}
\end{eqnarray}

\noindent and

\begin{eqnarray}
\eta(c \ge c_{win})= \eta_{cwwf}(c \ge c_{win}) +
\eta_{win}(c \ge c_{win}).
\label{(4)}
\end{eqnarray}

The   detailed  expressions for the wall and window frictions can
be found in ref. \cite{sierk2}.
\\
\\
\\
\noindent{\Large {\bf 3 Results}}
\\
\\
\noindent{\large {\bf 3.1 Fission widths from Langevin equation}}
\\
\\
With  all  the  necessary  input  defined  as above, the Langevin
equation (eq.(1)) is numerically integrated following the
procedure outlined in ref.\cite{ab1}. Starting with a given total
excitation energy ($E^{*}$) and angular  momentum  ($l$)  of  the
compound nucleus, the energy conservation in the following form,

\begin{eqnarray}
E^{*}=E_{int}+V(c)+p^{2}/2m
\label{(5)}
\end{eqnarray}

\noindent gives the intrinsic excitation energy $E_{int}$ and the
corresponding  nuclear  temperature $T=(E_{int}/a)^{1/2}$ at each
integration step. A Langevin trajectory  will  be  considered  as
having  undergone  fission  if  it  reaches  the  scission  point
($c_{sci}$) in the course of its time evolution. The calculations
are repeated for a large number (typically 100 000  or  more)  of
trajectories  and  the  number of fission events is recorded as a
function of time. Subsequently the time dependent fission   rates
can be easily evaluated.

\begin{figure}[htb]
\centering
\epsfig{figure=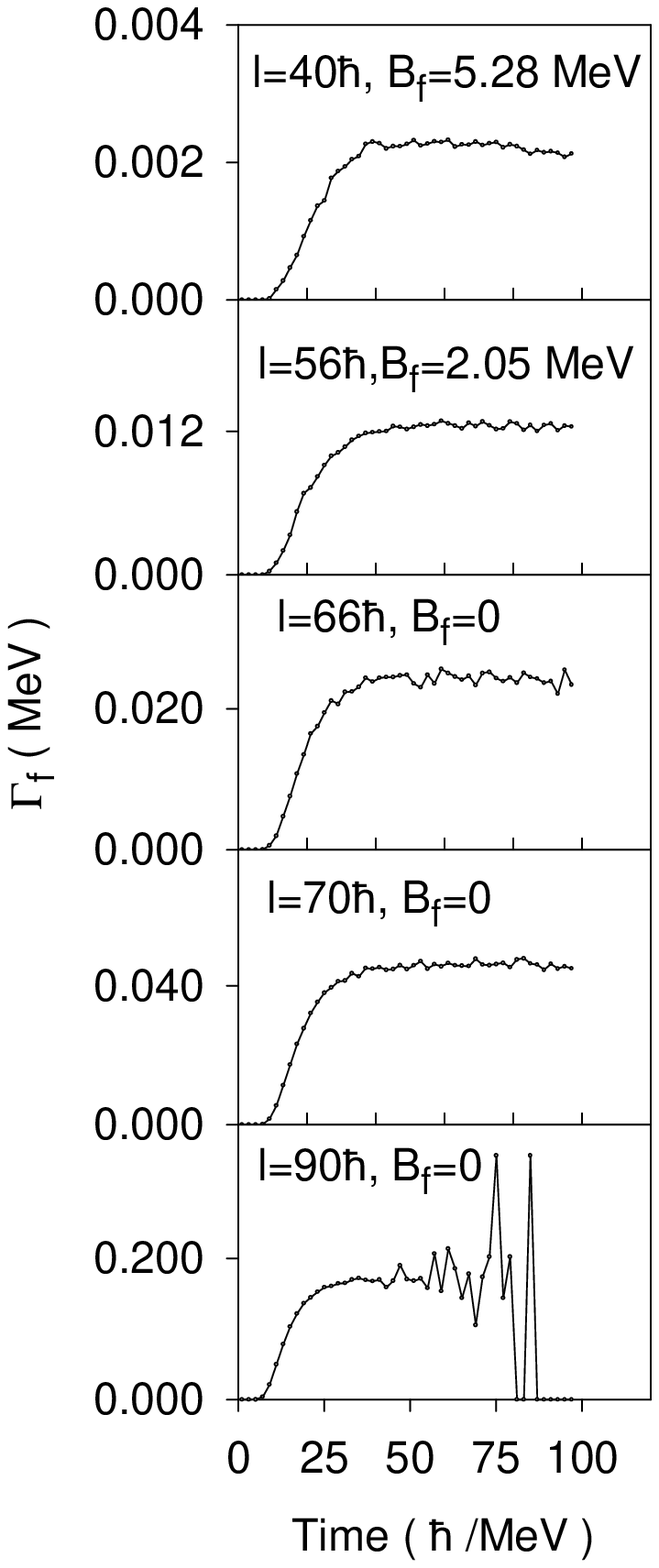}
\caption{\label{fig1}  Time   development   of   fission   widths
calculated  for  the compound nucleus $^{200}$Pb at a temperature
of    2  MeV  for  different nuclear spins $l$. The corresponding
values  of  the  fission  barriers  $B_{f}$  are   also   given.}
\end{figure}

We  have  chosen  the  $^{200}$Pb  compound nucleus for our study
which has  been  experimentally  formed  at  differnt  excitation
energies  in  a  number  of  heavy  ion  induced fusion reactions
\cite{newton,hinde1,hinde2}.   Figure   \ref{fig1}   shows    the
calculated  time  dependent  fission widths at different spins of
the compound  nucleus  for  a  given  temperature.  A  number  of
interesting  observations  can be made from this figure. The time
dependence of the fission width of the compound  nucleus  with  a
spin  of  40$\hbar$  (and with a fission barrier) is typical of a
diffusive flow across the fission barrier which has been  studied
extensively on an earlier occasion \cite{pal2}. The fission width
is  found  to  remain practically zero till a certain interval of
time ($t_{0}$) which  essentially  corresponds  to  the  interval
after  which  the  fission  trajectories  start  arriving  at the
scission point. The fission  width  subsequently  increases  with
time till it reaches its stationary value ($\Gamma_{0}$). We will
use  the following parametric form for the time dependent fission
width in order to enable us to use it in our later calculations,

\begin{eqnarray}
\Gamma  (t)  =  \Gamma_{0}   [1-   exp(-(t-t_{0})/\tau)] \Theta
(t-t_{0})
\label{(6)}
\end{eqnarray}

\noindent  where  $\tau$ is a measure of the transient time after
which  the  stationary flow is established and $\Theta(t)$ is the
step function. The intervals $t_{0}$ and $\tau$ are  obtained  by
fitting the calculated fission widths with the above expression.

We  next  note  in  fig. \ref{fig1} that the nature of the time
dependence of the fission width  remains  almost same even though
the fission barrier  decreases  and  subsequently  vanishes  with
increasing   spin.   At  very  large  values  of  spin,  however,
fluctuations appear at the later stages of time evolution.  These
fluctuations are statistical  in  nature  because  the  number of
nuclei which have not yet undergone fission decreases  very  fast
with  increasing  time  for  higher  values of spin and therefore
introduces large statistical errors in the measured numbers.  The
magnitude  of the fluctuations can thus be reduced by considering
a larger number of fission trajectories. In our  calculation,  we
have  taken  particular  care by using larger ensembles at higher
values of nuclear spin in order to enable us to check  whether  a
stationary value of the fission width is attained at all.

The  above  observation  is of particular interest since it shows
that the diffusive nature of  fission  persists  even  for  cases
which  have  no  fission  barrier.  This  diffusive  nature  is a
consequence  of  the  random  force   acting   on   the   fission
trajectories  as we have discussed earlier. As a compound nucleus
is formed having no potential pocket in the fission  channel,  it
starts  rolling  down  the  potential towards the scission point.
However, the random force acting on  these  fission  trajectories
introduces  a spread in their arrival time at the scission point.
The spread in the arrival time of the fission trajectories  gives
rise to a finite fission width as we find in fig. \ref{fig1}.

\begin{figure}[htb]
\centering
\epsfig{figure=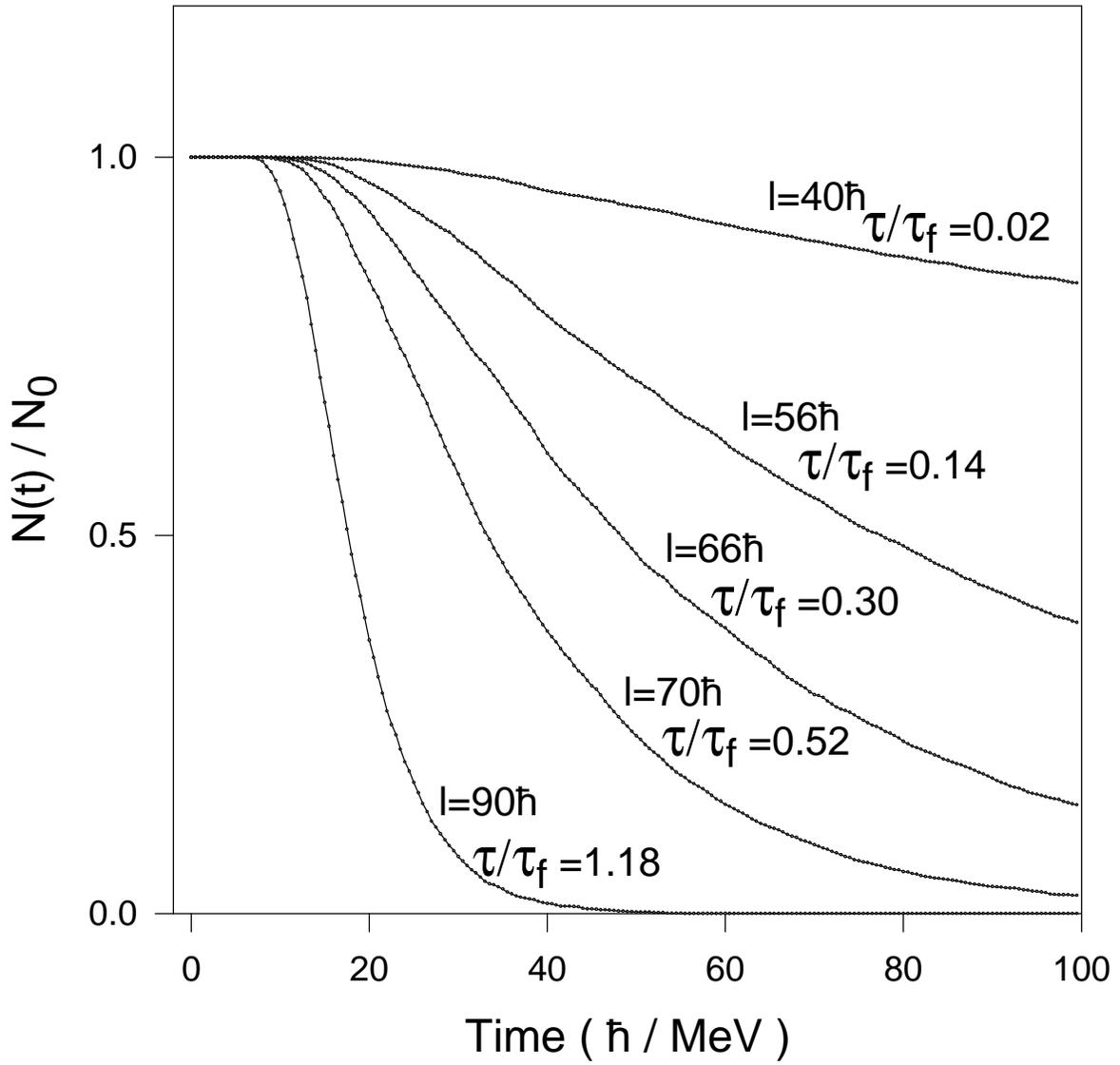}
\caption{\label{fig2}Survival probability of the compound nucleus
$^{200}$Pb against fission at a temperature of 2 MeV for different
nuclear  spins  $l$. The corresponding values of the ratio of the
transient  time to the fission life time ($\tau  / \tau_{f}$) are
also given.}
\end{figure}

\begin{figure}[htb]
\centering
\epsfig{figure=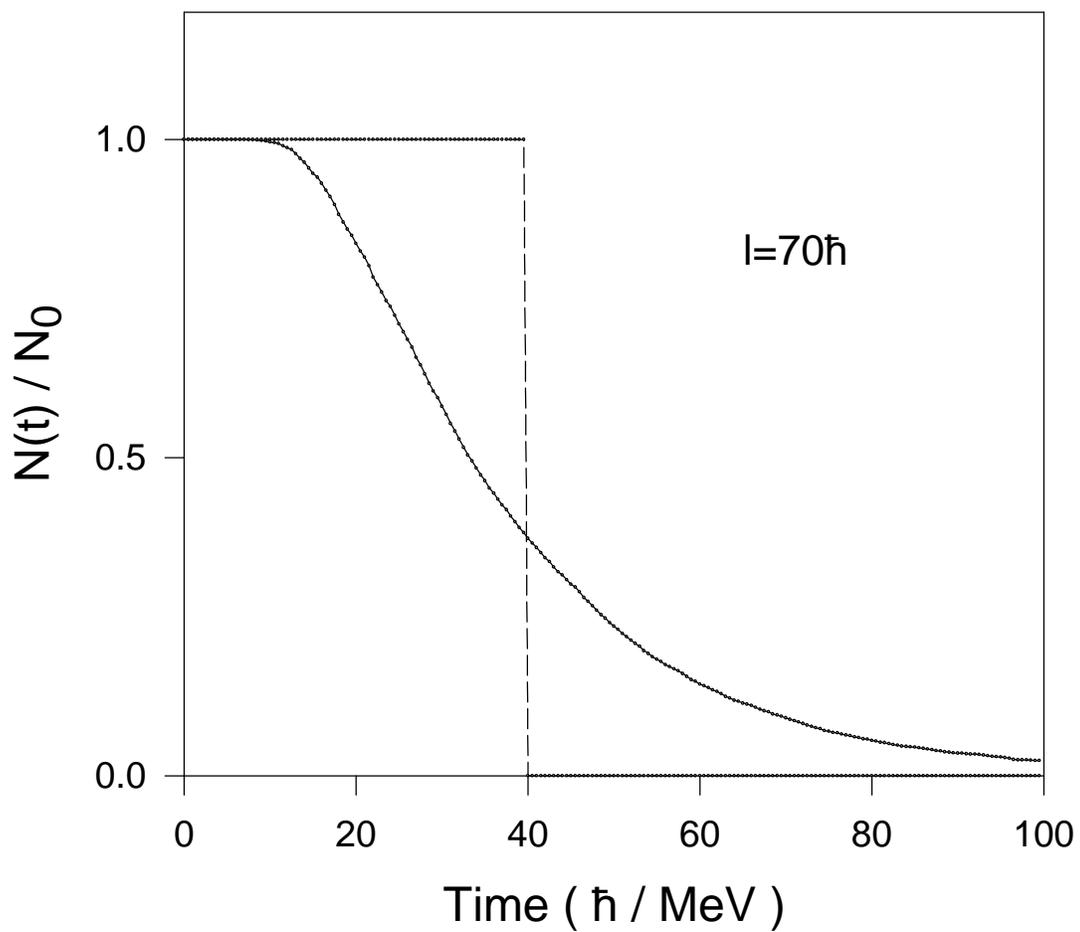}
\caption{\label{fig3}Survival probability of the compound nucleus
$^{200}$Pb  against  fission at a temperature of 2 MeV calculated
with (solid line) and without (dashed line) the random  force  in
the Langevin equation.}
\end{figure}

In  order  to  further  investigate the above diffusive nature of
fission, the fraction of the number of compound nuclei which have
survived  fission  is  shown  as  a  function  of  time  in  fig.
\ref{fig2}.  We  have  considered the same compound nuclei as in
fig. \ref{fig1} for this figure. Here we find a gradual shift  in
the  decay  rate  with  increasing  spin of the compound nucleus.
Specifically, the exponential decay of  the  number  of  compound
nuclei  having a fission barrier (with spins 40 and 56$\hbar$) is
found to continue for those without fission barriers (with  spins
66,  70  and  90$\hbar$).  Subsequently  we  have  calculated the
fraction of the  surviving  compound  nuclei  from  the  Langevin
dynamics  by  switching  off  the random force. Figure \ref{fig3}
shows  this decay in which all the nuclei have the same life time
which is simply the swooping  down  time  ($\tau_{s}$)  from  the
initial  to  the  scission  configuration. The spread in the life
time of the trajectories around this value when the random  force
is  switched  on  can also be seen in this figure. It may also be
noted that for very large values of the  compound  nuclear  spin,
the decay is very fast and consequently, the above spread is very
small. For such cases, fission is dominated by the transients and
can be approximated by a single swoop process.

\begin{figure}[htb]
\centering
\epsfig{figure=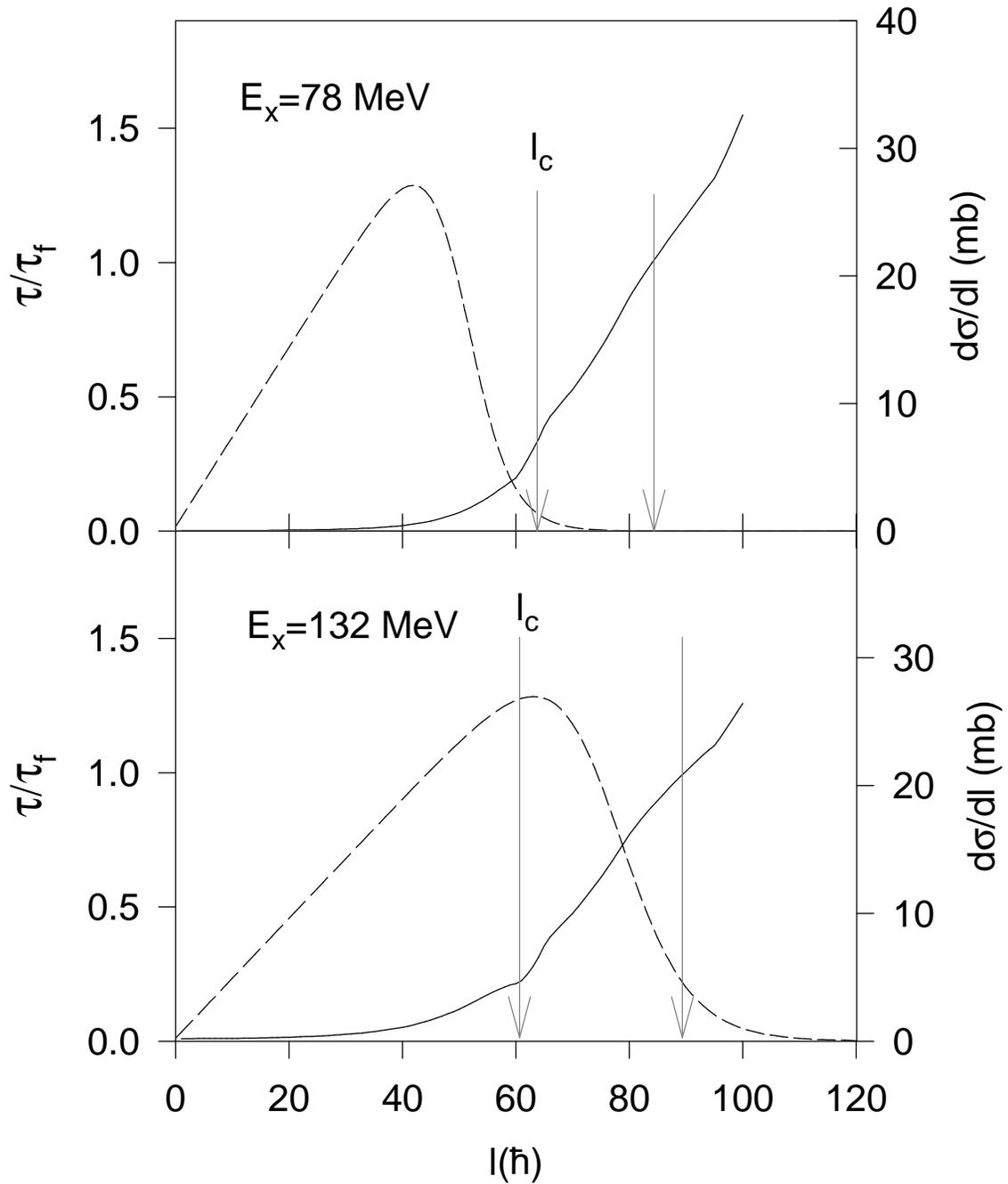}
\caption{\label{fig4}The  ratio  of  the  transient  time  to the
fission life time ($\tau / \tau_{f}$) as a function of  the  spin
$l$ of the compound nucleus $^{200}$Pb at two excitation energies
(solid  lines). The  transition  region  is  indicated by the two
arrows. The arrow at  the  critical  angular  momentum  ($l_{c}$)
marks  the  beginning  of  the  transition region. The next arrow
corresponds to $\tau / \tau_{f} =1$ and indicates the end of  the
transition  region.  The  partial  cross  sections  for  compound
nucleus formation are also shown (dashed lines).}
\end{figure}

We  shall  now  investigate  the  relevance of the different time
scales in order to distinguish between the  roles  of  stationary
flow  and  transients  in  fission.  When  the  fission life time
($\tau_{f}= \hbar/ \Gamma_{0}$) is much longer than the transient
time $\tau$, most of the fission  events  take  place  after  the
establishment  of  a  stationary  flow. Evidently, this holds for
nuclei with a barrier in the fission channel. However, it is also
possible to have $\tau_{f} > \tau$  for  cases  which  have  no
fission barrier. This is illustrated in fig. \ref{fig4} where the
ratio  $\tau  / \tau_{f}$ is plotted as a function of the spin of
the nucleus. Beyond the critical angular  momentum  $(l_{c})$  at
which  the  fission barrier vanishes, we find a window of angular
momentum where $\tau_{f}$ is indeed  greater  than  $\tau$.  This
window  represents  the  transition region over which the fission
dynamics changes  from  a  steady  flow  to  transients.  Fission
becomes   transient   dominated   for   spin   values   at  which
$\tau>\tau_{f}$. A single swoop description of  fission  can  be
applied  for such cases. However, a single swoop picture would be
rather inaccurate in the transition region where  a  steady  flow
still  persists.  In  the  next  subsection, we shall explore the
consequences of using the single swoop description of fission  in
statistical  model calculations in terms of the multiplicities of
prescission neutrons. It would  be  of  interest  for  our  later
discussions to locate the transition region with reference to the
spin  distributions  of  the  compound nuclei formed in heavy ion
induced  fusion  reactions.  We  have  therefore plotted the spin
distribution of the compound nucleus $^{200}$Pb obtained  in  the
fusion  of  $^{19}F$+$^{181}$Ta at two excitation energies. It is
observed that the transition region lies beyond the range of  the
spin distribution when the compound nucleus is excited to 78 MeV,
whereas  it is well within the range of the spin values populated
at an excitation of 132 MeV.  One  would  thus  expect  that  the
number  of prescisssion neutrons would be affected more at higher
excitation energies when the single swoop picture is used in  the
transition region.
\\
\\
\\
\noindent{\large \bf {3.2 Prescission  neutrons from dynamical and
statistical \linebreak  model calculation}}
\\
\\
We  shall  now consider the emission of prescission neutrons from
the Langevin dynamics of fission as well as  from  a  statistical
model  calculation  where  time-dependent  fission widths will be
used.  We  shall  consider  neutron  and  giant  dipole  $\gamma$
evaporation  in  the  Langevin  dynamical calculation following a
random sampling procedure \cite{fro3}. A Langevin trajectory will
be considered as undergone fission if  it  reaches  the  scission
point  in  course  of  its time evolution. Alternately it will be
counted  as  an  evaporation  residue  event  if  the   intrinsic
excitation energy becomes smaller than either the fission barrier
or  the  binding  energy  of  a neutron. The calculation proceeds
until the compound nucleus undergoes fission or  ends  up  as  an
evaporation  residue.  The number of emitted neutrons and photons
is recorded for each fission event. This calculation is  repeated
for  a  large  number  of  Langevin  trajectories and the average
number of neutrons emitted in the fission events  will  give  the
required prescission neutron multiplicity.

The statistical model calculation of prescission neutron emission
proceeds in a similar manner where a time-dependent fission width
is  used to decide whether the compound nucleus undergoes fission
in  each  interval  of  time  evolution. The intrinsic excitation
energy at each step is given by the total excitation energy minus
the rotational energy since no kinetic energy is associated  with
the fission degree of freedom in the statistical  model  and  the
compound   nucleus   is   assumed  to  be  in  its  ground  state
configuration   (zero   potenial   energy).   We  shall  use  two
prescriptions  for  the  time-dependent  fission  widths  in  our
calculation.  In  the first one, we shall use the parametric form
of  the  width  given  by  eq.(6)  for  all  spin  values
including  those  for  which  there  is  no  fission barrier. The
parameters $\Gamma_{0}$,  $t_{0}$  and  $\tau$  are  obtained  by
fitting  the numerically calculated time-dependent widths. In the
other statistical model  calculation,  we  shall  use  the  above
parametric  form  only  for  those spin values which have fission
barriers. For higher spin values for which there  is  no  fission
barrier including those in the transition region,  we  shall  use
the   swooping   down   picture.  For  these  cases,  we  shall
numerically obtain the swooping down time $\tau_{s}$ as explained
earlier. In  this  statistical  model  calculation,  neutron  and
$\gamma$ evaporation can take place during this period $\tau_{s}$
while  the nucleus will be considered as undergone fission at the
end of this interval.

\begin{figure}[htb]
\centering
\epsfig{figure=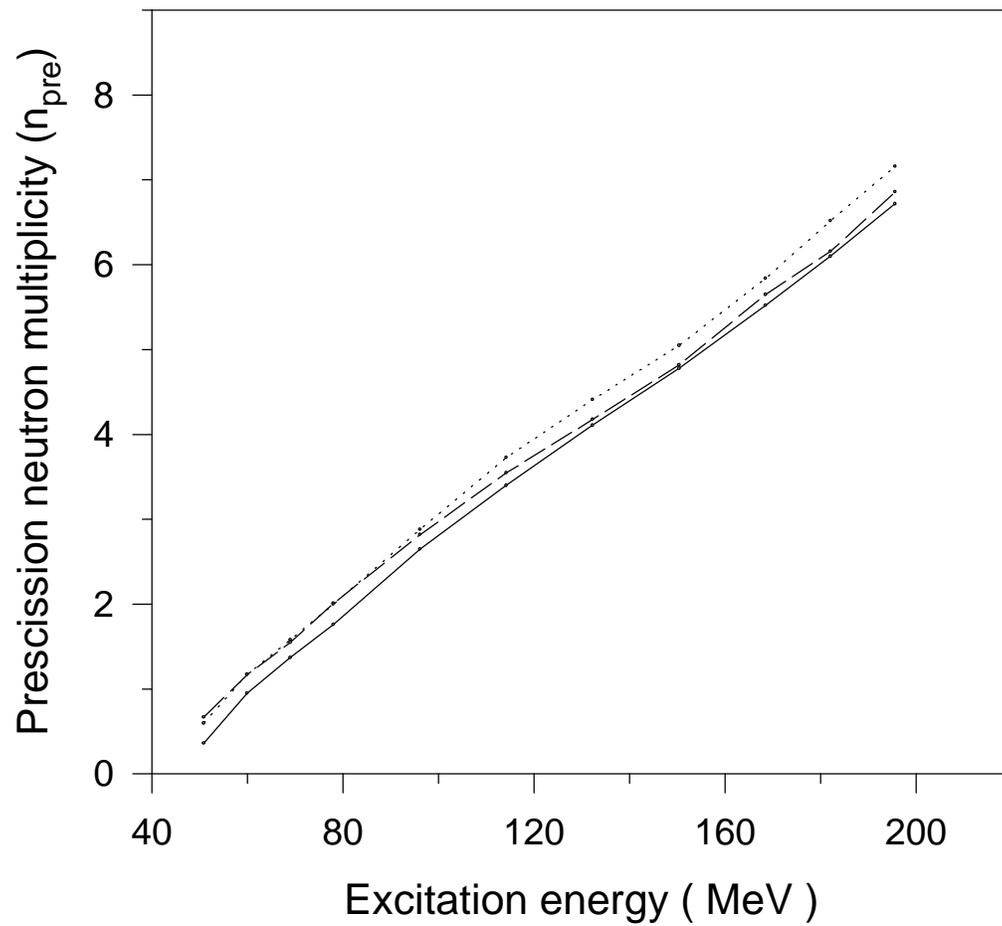}
\caption{\label{fig5}Prescission      neutron      multiplicities
calculated from the statistical  model  with  (dotted  line)  and
without  (dashed line) the single swoop approximation (see text).
Results from the dynamical model (solid line) are also shown.}
\end{figure}

Figure   \ref{fig5}  shows  the  calculated  prescission  neutron
multiplicity at different excitation  energies  of  the  compound
nucleus $^{200}$Pb  formed in the $^{19}$F + $^{181}$Ta reaction.
Results shown in this figure are obtained from the dynamical  and
statistical  model  calculations which are continued for a period
of 300$\hbar/MeV$. This time period is not sufficient for all the
nuclei in the ensemble either to reach the  fission  fate  or  to
become  evaporation  residues.  Pushing  the Langevin calculation
much beyond the above time period becomes prohibitive in terms of
computer time. The above time duration  is  however  much  longer
than  the  transient times and hence are adequate for our purpose
of comparing the dynamical and statistical results.

\begin{figure}[htb]
\centering
\epsfig{figure=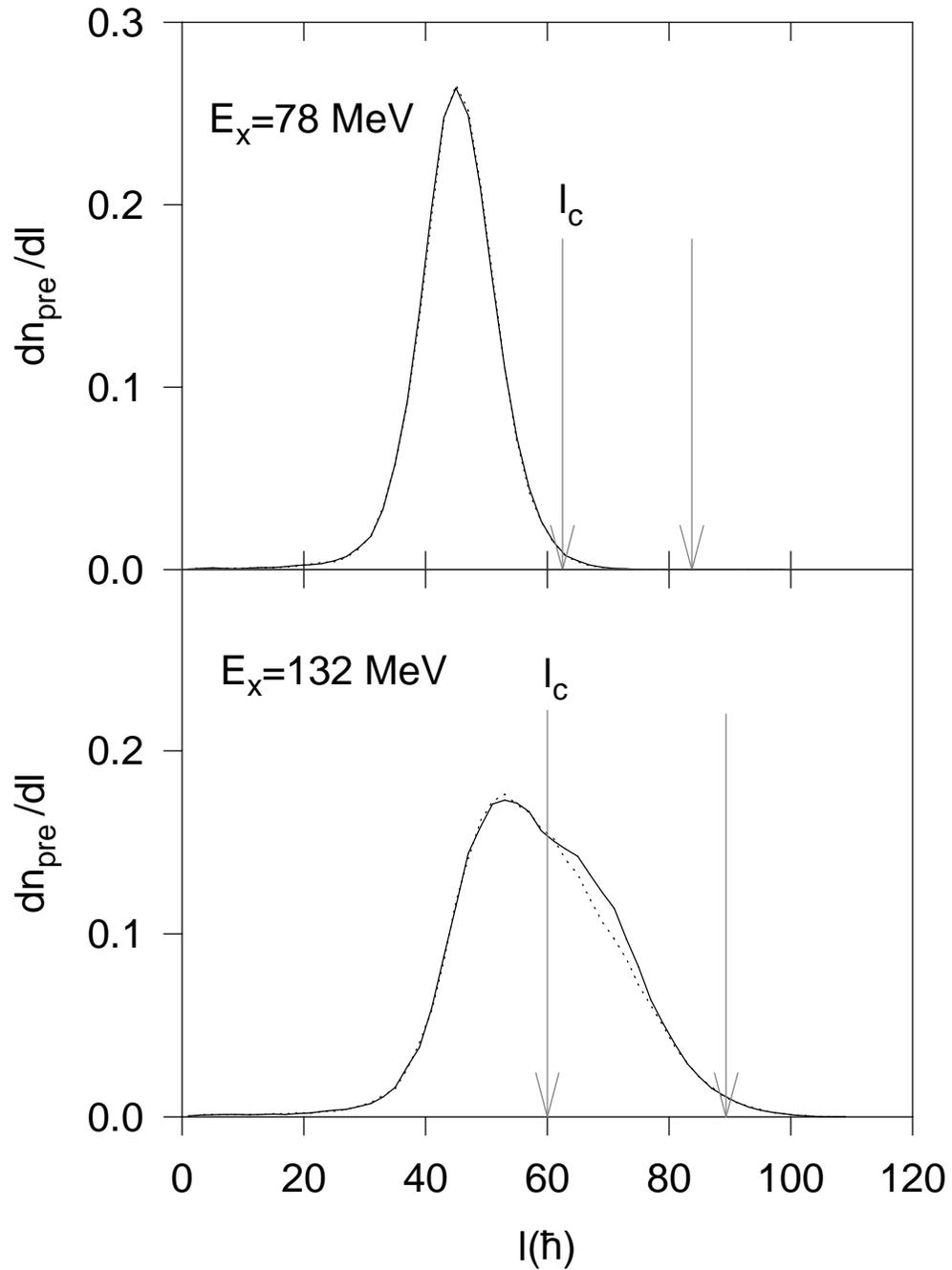}
\caption{\label{fig6}
Differential prescission neutron multiplicities  calculated  with
the single swoop approximation at two excitation energies  (solid
lines). The corresponding distributions  without the single swoop
approximation are shown  by  the  dotted  lines.  The  transition
regions are also indicated as in fig.4.}
\end{figure}

We first note in fig.\ref{fig5}  that  the  neutron  multiplicity
calculated from the statistical model  using  the  time-dependent
fission  widths  with  and  without  swooping down assumption are
alomost same at lower  excitation  energies  though  they  differ
marginally  at  higher excitation energies. Such a difference was
anticipated in the earlier subsection  since  the  swooping  down
assumption is invoked more frequently for compound nuclei at high
excitation  energies which are mostly formed with large values of
spin and consequenlty  with  no  fission  barrier.  In  order  to
explore    this   point   further,   the   differential   neutron
multiplicities  are   obtained   from   the   statistical   model
calculations with as well as without the single swoop description
and   are   shown   in   fig.   \ref{fig6}.  The  two  calculated
distributions  at an excitation energy of 132 MeV are found to be
different beyond $l_{c}$ though they merge again  at  the  higher
end   of  the  transition  region.  This  difference  essentially
reflects the approximate nature of the single  swoop  description
in   the  transition  region.  However,  the  magnitude  of  this
difference is found to be rather small ($\sim$ a few $\%$). At  a
lower  excitation  of  78  MeV,  the two distributions are almost
identical as one would expect since they have very little overlap
with  the  transition  region.  The  significance  of  the  above
observations is of interest since  it  shows  that  for  compound
nuclei without a fission barrier, considering a sharp valued life
time  (the  swooping down time $\tau_{s}$) instead of a life time
with a dispersion does not make any  appreciable  effect  in  the
number of emitted neutrons before fission. It is next observed in
fig.\ref{fig5} that the neutron multiplicity from the statistical
(both  calculations)  and dynamical models are also very close to
each other though the statistical models marginally  overestimate
the  neutron  multiplicity  compared  to  the  dynamical model. A
possible explanation for this observation would be the fact  that
the  compound  nuclear  temperature  in  the statistical model is
higher than that in the dynamical model since a part of the total
excitation energy is locked up as kinetic energy of  the  fission
mode   in   the  dynamical  model.  This  reduces  the  intrinsic
excitation energy and hence  the  temperature  in  the  dynamical
model resulting in a smaller number of evaporated neutrons.

We  have  already mentioned that a full dynamical calculation can
take an extremely  long  computer  time  particularly  for  those
compound  nuclei  whose  fission  probability  is small. We shall
therefore follow a  combined  dynamical  and  statistical  model,
first  proposed by Mavlitov {\it et al.} \cite{fro2}, in order to
perform a full calculation. In this model, we shall first  follow
the  time  evolution  of  a  compound  nucleus  according  to the
Langevin equations for a sufficiently long period  during which a
steady flow across the fission barrier is established.  We  shall
then  switch  over  to  a statistical model description after the
fission process reaches the stationary regime. It is possible  to
continue  this calculation for a sufficiently long time such that
every  compound  nucleus  can  be  accounted  for  either  as  an
evaporation residue or having undergone fission.

\begin{figure}[htb]
\centering
\epsfig{figure=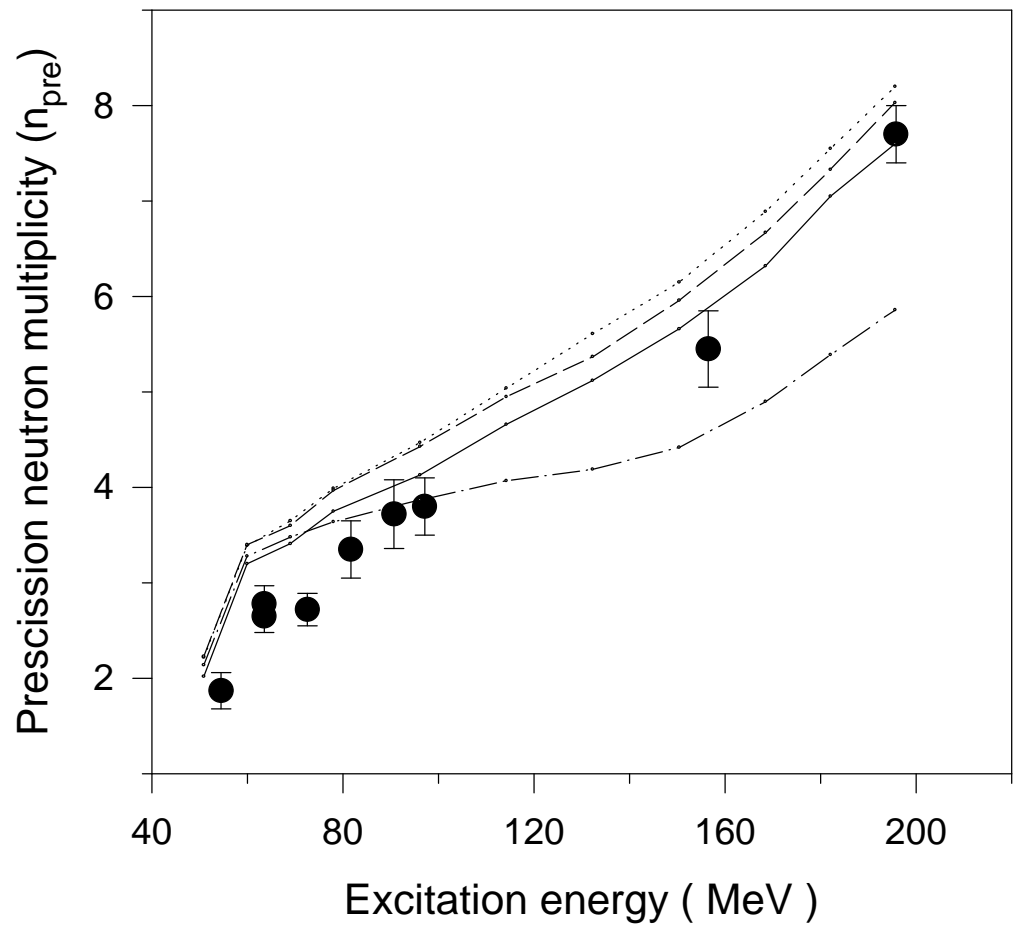}
\caption{\label{fig7}
Prescission    neutron   multiplicities   calculated   from   the
statistical model with (dotted line) and  without  (dashed  line)
the  single swoop approximation and also from the dynamical model
(solid line) along with the experimental  data.  The results of a
statistical  calculation  using  the  stationary  values  of  the
fission widths are also shown (dash-dotted line). }
\end{figure}

The  prescission  neutron  multiplicity calculated with the above
combined  dynamical  and  statistical  model  is  shown  in  fig.
\ref{fig7}  along  with  the full statistical model calculations.
The  statistical  model  calculations  are  made  with as well as
without the swooping down assumption in  the  time-dependence  of
the  fission  widths.  The  experimental values are also shown in
this figure. The observations made in this figure are similar  to
those  in  fig.\ref{fig5},  namely,  the statistical calculations
slightly overestimate the neutron multiplicity  compared  to  the
dynamical    (plus   statistical)   calculation.   However,   the
statistical and dynamical results are quite close to  each  other
and  are  also  close  to  the  experimental  values. This result
therefore   shows   that   the   statistical   calculation   with
time-dependent   fission   width   can  represent  the  dynamical
calculation with reasonable accuracy.  We  have  also  shown  the
results  of  a  statistical  calculation in this figure where the
fission widths are assumed to be  independent  of  time  and  are
given  by their stationary values. This calculation substantially
underestimates  the  neutron  multiplicity  and  illustrates  the
importance of transients at higher excitation energies.
\\
\\
\\
\noindent{\Large {\bf 4 Summary and conclusions}}
\\
\\
We  have  presented  in  the  above  a  numerical  study  of  the
transients in the fission of  highly  excited  nuclei  and  their
effect  on  the  number  of neutrons emitted prior to fission. To
this end, we first investigated the  time-dependence  of  fission
widths using the Langevin dynamics of fission. We have shown that
the  fission  width  reaches a stationary value after a transient
period even for those nuclei which have no  fission  barrier.  We
have discussed the role of the random force acting on the fission
trajectories in introducing a dispersion in their arrival time at
the  scission  point  and thereby giving rise to a finite rate of
fission for such cases. We have also shown that  this  stationary
fission  rate  for very large values of spin of the nucleus loses
significance  since  the  stationary  fission  life  time  itself
becomes  much  smaller  than  the  transient time for such cases.
Therefore, fission  of  nuclei  rotating  with  a  large  angular
momentum  can  be  considered  to  proceed in a single swoop. Our
study demonstrates a gradual transition from  a  diffusive  to  a
single  swoop  picture  of  fission  with  increasing spin of the
compound nucleus.

We have subsequently examined the effect of the transients on the
multiplicity  of  the  prescission  neutrons emitted in heavy ion
induced  fusion-fission  reactions.  We  used  both the diffusive
description  and  the  swooping  down   picture   separately   in
statistical  model calculations and found close agreement between
the two calculated neutron numbers  at  low  excitation  energies
whereas  they  differed  marginally at higher excitations. It was
also  shown  that   the   differential   neutron   multiplicities
calculated  with  and  without the single swoop assumption differ
only in  the  transition  region  though  the  magnitude  of  the
difference is small. We therefore conclude that  the single swoop
description  of  fission  can  be  used  in   statistical   model
calculations without making any  significant  error  in the final
observables.

We  finally  compared    the number of neutrons calculated from a
dynamical model with that obtained from a  statistical  model  in
which  time-dependent  fission widths are used. We found that the
statistical model marginally overestimates  the  neutron  numbers
than those from the  dynamical  calculation.  We  explained  this
difference  in  terms  of  the  temperature which is lower in the
dynamical model than the statistical calculation. The temperature
turns out to be  smaller  in  the  dynamical  model  because  the
excitation  energy  is shared between the collective fission mode
and the thermal mode in the dynamical calculation in contrast  to
the  statistical  calculation where the full excitation energy is
assumed to be available in the thermal mode. However, in most  of
the  fission  events  in  the  dynamical calculation, the kinetic
energy builds up to values which are a little above  the  fission
barrier  before  it  proceeds to fission. Since the values of the
fission barrier (typically a few MeV or less)  are  much  smaller
than  the  excitation  energies  (a  few  tens  of  MeV  or more)
considered  here,  the  temperature   differences   between   the
statistical  and  dynamical calculations remain small for most of
the cases. Consequently the difference  between  the  prescission
neutron   multiplicities   calculated   from  the  dynamical  and
statistical models become small,  as  we  have  observed  in  our
calculation.

\eject

\end{document}